\documentstyle[12pt]{article}
\begin{document}
\thispagestyle{empty}
\begin{center}
{\LARGE \tt \bf Torsion Gravity Effects on Charged-Particle and Neutron 
Interferometers}
\end{center}

\vspace{1cm}

\begin{center}
by C. Sivaram and L.C. Garcia de Andrade\footnote{Departamento de F\'{\i}sica Te\'{o}rica 
- Instituto de F\'{\i}sica -UERJ - Rua S\~{a}o Francisco Xavier, 524, 
Maracan\~{a} RJ - Rio, Brasil.

\vspace{2cm}

.}
\end{center}

\begin{abstract}
Torsion gravitational effects in the quantum  interference of charged 
particles are investigated. The influence  of  axial torsion  in  the 
Schiff-Banhill effect (SB) inside a metallic shell is given. The effect
of torsion on the surface of the earth on (SB) experiment is estimated.
Torsion  gravity  effects on  the Sagnac phase-shift of neutron 
interferometry are also computed.

\end{abstract}
\newpage

Earlier, Anandan \cite{1}  have consider neutron interferometer experiments 
in spaces with torsion. In his analysis he arrived at conclusion that 
to produce a phase-shift of  $\Delta \theta \cong 10^{-3}$ , as in current  
experiment  with thermal neutrons beans, a torsion field of  $Q\cong 10^{-9}cm^{-1}$ would be needed. Anandan \cite{2} also showed that in the Einstein-Cartan of Spin and Torsion these experiments  of  Collela, Overhauser  and Werner (COW) type\cite{3,4} would require a very weak torsion field of $Q\cong 10^{-43}cm^{-1}$. Just to give an idea of how weaker this 
field is the torsion field  on  the surface  of the Earth is 
$Q_{\star}\cong 10^{-24}s^{-1}$. Later Anandan \cite{5} considered the (SB) 
experiment\cite{6} in the gravitational field of the earth. In this paper,  
I consider the charged-particle interferometry (SB) experiment on torsion 
backgrounds. The basic new feature here is  that  the torsion  field of 
the Earth is the one given by Nitsch and Hehl\cite{7}; (see also 
Nitsch\cite{8}) using a PPN aproximation of a translational gauge theory of 
gravity with torsion. Their result is  $Q_{\star}\cong 10^{-24}s^{-1}$
Unfortunately, the torsion  effects  on  the phase-shift are small as in 
the Aharonov-Bohm (A-B) case\cite{9,10} but are interesting from the 
theoretical point of view. Nevertheless the Sagnac \cite{11} phase-shift of 
the Earth rotation on neutron interferometry yields a very interesting 
application of torsion since Nitsch-Hehl\cite{7} formula contain a relation 
between the  rotation  of the astrophysical objects (planets, stars)  and 
torsion. Which yields a straight forward torsion contribution to the Sagnac 
effect. 

Let us now first consider  the  extension  of SB effect that there must 
exist an electric field $\vec{E_{S}}$ inside  a  metallic  shell that has no
currents. This  $\vec{E_{S}}$ satisfies  $m\vec{g} + e\vec{E_{S}} = \vec{0}$ 
in the nonrelativistic limit.

The  SB  idea  was  proved\cite{12}  inside  a  hollow  cylinder  only  for 
temperatures of ~<4.2K. Above this temperature, this  field  undergoes
dramatic changes that are not yet understood.

Let us consider the extension of the SB equation to include torsion as:
\begin{equation}
0 = m\frac{d\bar{v}^{\mu}}{dT} = -\left\{{\nu}^{\mu}\ \rho \right\}
\bar{v}^{\nu}\bar{v}^{\rho} + F^{\mu}_{\nu}\bar{v}^{\nu}- \frac{3}{4}S^{\nu}
{\partial }_{\nu} Q^{\mu}
\label{1}
\end{equation}

Where $v^{\mu}\cong \hbar k^{\mu}/m\ , k^{\mu}$ being the wave vector, for a  
typical electron and the bar denotes averaging over the  3-velocity  u and 
neglecting  $O\left( u^{4}/c^{4} \right)$ terms. The last therm  in  equation 
(\ref{1}) has been computed by Sabbata and Gasperini \cite{13}. Writing 
equation (\ref{1}) in  3-vector rotation yields.

\begin{equation}
m\vec{g} + e\vec{E_{S}} - \frac{3}{4}( \vec{S}. \nabla )\vec{Q} = \vec{0}
\label{2}
\end{equation}

Which is the generalization of SB  equation to include torsion effects.
Just considerating torsion contributions ans electrical effects, we are
left with:
\begin{equation}
\Delta {\theta}_{SB} = - \frac{e}{\hbar}\ \delta A_{0} T = 
+ \frac{e}{\hbar}\left(\int \vec{E_{S}}. \ d\vec{r} \right)T
\label{3}
\end{equation}

Which implies for the phase-shift torsion contribution.
\begin{equation}
\Delta {{\theta}_{SB}}^{torsion}= \frac{Qd}{v} = \frac{mQA}{r\hbar K}
\label{4}
\end{equation}

Where r is the distance between two hollow cylinders in SB experiment.
And $v \cong 10^{8}cm/s$ for electrons in copper. Considerating the torsion 
field contribution from the gravitational field of the earth 
$Q_{\star}\cong 10^{-24}s^{-1}$ this effect is extremelly weak.
\begin{equation}
\Delta {{\theta}_{SB}}^{torsion} = 10^{-19}d
\label{5}
\end{equation}

In the nonrelativistic limit of the torsion less case 
$\Delta {\theta}_{SB} = - \Delta {\phi}_{cow}$ where $\Delta {\phi}_{cow}$
is the phase-shift of the Collela, Overhauser, Werner (COW) experiment
given  by $ \Delta {\phi}_{cow} = - \frac{m^{2}g A}{{\hbar}^{2} K}$ In our 
case the torsion contribution to the Cow experiment is extremely weak and 
does not affect the experiment. To see this is enough to check that 
$\Delta {{\theta}_{SB}}^{torsion}/\Delta {\theta}_{SB} \cong 
\frac{Q}{mr} \cong 10^{-24} \cong 10^{26} !$ which is a very small number to 
introduce any measurable results on the experiment. Despite of this null 
result for SB experiment  I  shall  demonstrate next that the influence of 
torsion on the Sagnac effect is not so small.

Let us consider a torsion contribution to the Sagnac effect\cite{14}. This
effect  is  very  similar  to  the  London  moment  equation  for  a 
superconductor with an angular velocity $\vec{\Omega}$\cite{5}.
\begin{equation}
e\vec{B_{L}} + 2m \vec{\Omega} = 0
\label{6}
\end{equation}

The Sagnac effect yields a phase-shift as:
\begin{equation}
\Delta {\phi}_{S} = - \frac{2m}{\hbar}\int \vec{\Omega}\ . d\vec{S}
\label{7}
\end{equation}

Where S is a surface spanned  by  the neutron beans in the neutron 
interferometry experiment. $\vec{\Omega}$ is the earth's rotation in the case  
of terrestial experiments. By inverting the Nitsch-Hehl formula one obtains a 
relation for $\vec{\Omega}$ in therms of torsion.
\begin{equation}
\vec{\Omega} = \left( \frac{R}{R_{S}}\right)^{2}\vec{Q}
\label{8}
\end{equation}

Where $R_{S} = \frac{2GM}{c^{2}}$ is the Schwaszschild radius.
Substituion of (\ref{8}) into (\ref{7}) one obtains a Sagnac phase-shift due  
to torsion.
\begin{equation}
\Delta {{\phi}_{S}}^{torsion}= - \frac{2mQ}{\hbar}\left(\frac{R}{R_{S}} 
\right)^{2}A
\label{9}
\end{equation}

In the case of the earth, $Q_{\star}\cong 10^{-24}s^{-1}\ , m_{n}
\cong 10^{-24}g\ , \hbar \cong 10^{-27}cgs\ units$ , $R_{S}\cong 10^{6}cm$
and $R_{\star}\cong 10^{8}cm$. Substitution of these data into (\ref{9}) 
yields.
\begin{equation}
|\Delta {{\phi}_{S}}^{torsion}| \cong 10^{-17}A
\label{10}
\end{equation}

This gives some hope to detect torsion increasing the $\alpha $ area enclosed
by the neutron beams, for terrestial laboratories experiments a typical
value for the area is $A \cong 10^{4}cm^{2}$ , for this value the 
Sagnac-torsion effect is $\Delta {{\phi}_{S}}^{torsion}\cong 10^{-13}$ which 
is a small value to be detected. This is of the same order of the 
Aharonov-Bohm (A-B) effect on iron magnet rotating tube.

For example the A-B effect computed by H.Peng \cite{14} the phase-shift is
given by:
\begin{equation}
\Delta {\phi}_{A-B} = \frac{4m}{\hbar}\int \int \vec{B_{g}}\ . d\vec{S} =
\frac{2Gm\pi R^{4}\rho}{\hbar c^{2}}\ \Omega = 10^{6}m
\label{11}
\end{equation}

Where m is the particle's mass and $\vec{B_{g}}$ is the gravitomagnetic field. 
For electrons $\Delta {\phi}_{A-B}\cong 10^{-21}$ which is much weaker than 
the torsion  contribution to the Sagnac phase-shift. More on this can be 
found on a recent paper by Garcia de Andrade and Sivaram on the Torsion 
Gravitational A-B Effect\cite{10}.

A more recent account on the Sagnac effect  on  neutron interferometry
has been considered recently by B.Mashoom  in the framework of Special
Relativity. A more detailed investigation  on the torsion influence on
the Sagnac effect can appear elsewhere.

\section*{Acknowledgements}

I would  like  to  express  my  gratitude  to  Professors J.Anandan,
C.M.Zhang, and Dr. H.Peng  for  helpful discussions  on  the 
subject of this paper. Finantial support from  CNPq (Conselho Nacional
de Pesquisas, Brasil) is gratefully acknowledged.

\end{document}